\documentclass[conference]{IEEEtran}
\IEEEoverridecommandlockouts

\usepackage{cite}
\usepackage{amsmath,amssymb,amsfonts}
\usepackage{algorithmic}
\usepackage{graphicx}
\usepackage{textcomp}
\usepackage{xcolor}
\usepackage{tikz}
\usepackage[version=4]{mhchem}
\usepackage{booktabs}
\usepackage{multirow}
\usepackage{siunitx}
\usepackage{eurosym}
\usepackage{mhchem}
\usepackage{amssymb}
\usepackage{tikz}
\usepackage{pgfplots}
\usepackage{multirow}
\usepackage{graphicx}
\usepackage{hyperref}
\usepackage{eurosym}
\def\BibTeX{{\rm B\kern-.05em{\sc i\kern-.025em b}\kern-.08em
    T\kern-.1667em\lower.7ex\hbox{E}\kern-.125emX}}
\usepackage{url}

\begin{document}

\title{Synergy Among Flexible Demands: Forming a Coalition to Earn More from Reserve Market}

\author{Peter A.V. Gade\textsuperscript{*}\textsuperscript{\textdagger}, Trygve Skjøtskift\textsuperscript{\textdagger}, Henrik W. Bindner\textsuperscript{*}, Jalal Kazempour\textsuperscript{*} \\
    \textsuperscript{*}Department of Wind and Energy Systems, Technical University of Denmark, Kgs. Lyngby, Denmark \\
    \textsuperscript{\textdagger}IBM Client Innovation Center, Copenhagen, Denmark
    \thanks{
        Email addresses: pega@dtu.dk (P.A.V. Gade), Trygve.Skjotskift@ibm.com (T. Skjøtskift), hwbi@dtu.dk (H.W. Bindner), jalal@dtu.dk (J. Kazempour).}

    \vspace{-3mm}
}




\maketitle


\IEEEaftertitletext{\vspace{-0.8\baselineskip}}
\maketitle
\thispagestyle{plain}
\pagestyle{plain}
\begin{abstract}
    We address potential synergy among flexible demands and how they may earn more collectively than individually by forming a coalition and bidding to the reserve market.
    We consider frequency-supporting ancillary service markets, particularly the manual Frequency Restoration Reserve (mFRR) market.
    The coalition of flexible demands provides more reliable mFRR services, where in comparison to individual demands, is penalized less for their potential failure and is paid more for their successful activation. This synergy effect is quantified as a function of the number of homogeneous assets in the coalition. A subsequent payment allocation mechanism using Shapley values is proposed to distribute the total earnings of the coalition among demands, while incentivizing them to remain in the coalition. For our numerical study, we use real price data from the Danish mFRR market in 2022.
\end{abstract}



\vspace{-2mm}
\section{Introduction}\label{sec:Introduction}
\subsection{Background}
It has been extensively addressed in the literature how the aggregation of flexible demands can provide frequency-supporting ancillary services to the power system \cite{biegel2014value}, \cite{macdonald2020demand}. By ancillary services, we refer to various forms of reserves, booked by the Transmission System Operator (TSO) in advance, to activate them in the operational stage, if necessary. The key for ancillary service provision by the aggregation of flexible demands is the availability of a \textit{baseline forecast} of total consumption. The TSO should approve in advance the process adopted by the aggregation of demands for their baseline forecast. This has been prescribed in the official documentation of the Danish TSO, Energinet, issued for the pre-qualification of demand-side resources for ancillary service provision \cite{energinet:prequalification}.

The aggregation of flexible demands is incentivized for at least two reasons: (\textit{i}) to meet the minimum bid size requirement, if it exists, for ancillary service markets \cite{energinet:Systemydelser}, and (\textit{ii}) to unlock the potential \textit{synergy} among flexible demands. By synergy, we refer to operational circumstances under which the aggregation of flexible demands, compared to individual demands, may provide ancillary services with a higher level of reliability, i.e., a lower rate of delivery failure in the activation stage. By this, the whole aggregation is penalized less for their potential failure and paid more for their successful activation. We do not consider a case that, by aggregation, flexible demands may offer a larger volume of ancillary services. Hereafter, we refer to the aggregation of flexible demands as a \textit{coalition}, being operated by an \textit{aggregator}. Here, we consider that the aggregator represents the coalition, and may charge flexible demands for its service, i.e., facilitating the creation of the coalition, which is not part of this study.

Several private initiatives for the provision of demand-side flexibility via forming coalitions are currently taking place in Denmark. For example, IBM is developing their \textit{Flex Platform}, aiming to aggregate and control heterogeneous assets in commercial and industrial sectors. The aggregated flexibility is then shaped as flexibility services to be sold in various ancillary service markets running by the TSO, among which the market for manual Frequency Restoration Reserve (mFRR), also known as tertiary reserve, has usually the largest trading volume. The collected profit is supposed to be allocated among flexible demands in a fair manner. In addition, flexible demands might be further incentivized by being informed about their contribution to the CO$_{2}$ savings.

\subsection{Research questions and our contributions}
\vspace{-1mm}
Aggregation platforms such as the \textit{Flex Platform} of IBM face two fundamental questions: (\textit{i}) How many demand-side assets are needed to be aggregated in order to unlock the potential synergy, resulting in sufficiently reliable ancillary services as prescribed by the TSO? and (\textit{ii}) How to allocate payments ex-post to individual flexible demands within the coalition?

We answer the first question by quantifying the synergy effect in terms of total earning of the coalition as a function of the number of assets within the coalition. For simplicity, we assume identical flexible assets, and limit our question to their number. The future work should extend the question by also considering the heterogeneity of assets as a further degree of freedom. We show how individual assets are inherently unpredictable on their own, but comparatively more predictable within the coalition.

We then answer the second question by proposing Shapley values \cite{shapley1997value} as a mechanism to reward flexible demands for their contribution to the coalition. This provides desirable economic properties such as individual rationality and budget balance that incentivize flexible demands to stay in the coalition, as opposed to acting individually. We illustrate this payment mechanism using a stylized example of five flexible demands, each with numerous assets, in a coalition, for which the aggregator bids in the mFRR market only. We conduct our simulations based on real price data of the Danish mFRR market from 2022.






\vspace{1mm}
\subsection{Status quo}
\vspace{-1mm}
Numerous studies in the literature explored important aspects of demand-side flexibility such as feasibility and controllability \cite{bondy2018redefining}, \cite{bondy2017performance}, \cite{bondy2016procedure}, \cite{bondy2014performance}, \cite{biegel2014integration}, \cite{AchievingControllabilityofElectricLoads}. However, an important aspect that has been largely overlooked is the potential synergy among demand-side assets, such as freezers, heat pumps, ventilation systems, etc.
Reference \cite{biegel2014value} investigates the value of flexible consumption but do not touch upon the synergy effect of a portfolio of assets, although it is mentioned that market barriers exist such as the minimum bidding size.
It has been shown in \cite{pedersen2014aggregation} how a portfolio of supermarkets can deliver a granular power response which is made possible by the number of supermarkets and their assets. Although it has not been explicitly mentioned, it refers to a form of synergy effect, because an aggregator requires a certain number of controllable assets to deliver a granular power response. However, the estimation and value of flexibility ex-ante is not tied together with any synergy effect. Note that demand-side synergy is also known in other domains, e.g., for strategic and organizational diversification \cite{ye2012achieving}, but here we specifically address demand-side flexibility for ancillary services in power systems.

For the coalition of flexible demands, the baseline estimation is key, from which the aggregator estimates the flexible capacity, as prescribed by the Danish TSO \cite{energinet:prequalification}. Note that the coalition of flexible demands does not necessarily have an operational baseline schedule as a generation unit does \cite{gade2022ecosystem}.
In this work, we show that the demand-side flexibility provision becomes slightly less sensitive to the accuracy of the baseline estimation due to the synergy effect among assets within the coalition.
In addition, it is important to estimate the baseline while being aware of potential counterfactual consumption when a flexible demand is activated. This problem has been studied extensively. For example, \cite{ziras2021baselines} explains why baselines are not necessarily suited for harnessing flexibility in the distribution grid, whereas \cite{capacity_limitation_services} instead shows how a mechanism based on capacity limitation services may work more successfully in practice. Nonetheless, for the TSO-level ancillary services aiming to instantaneously balance total production and total demand in the entire system, the baseline approach is generally adopted as a common practice. Some studies have proposed more innovative mechanisms as alternatives. For example, \cite{muthirayan2019mechanism} proposes a mechanism by which the aggregator relies on self-reported baselines from individual flexible demands, removing the incentive to inflate baselines. In our setting, this is not straightforward to be implemented, since flexible demands have generally no expertise or even motivation to report their own baselines. Recall that flexible demands have their other primary business purposes, which is not flexibility provision, and prefer to not get involved in complicated mechanisms.



\vspace{0mm}
\subsection{Paper organization}
\vspace{-1mm}
The rest of the paper is organized as follows.
Section \ref{chapter2} introduces the proposed simulation setup. This includes an introduction to the mFRR services, the mathematical expression of the coalition profit and the synergy effect, and finally the description of the Shapley values as the payment mechanism. Section \ref{chapter3} provides simulation results, illustrating the synergy among a number of assets and the corresponding payment allocation mechanism. Lastly, Section \ref{chapter4} concludes the paper and outlines potential directions for the future work.

\section{Simulation setup}
\label{chapter2}

\subsection{mFRR services}\label{sec:mFRR}
Ancillary services including mFRR are exploited by the TSO to balance the power system such that at any instant the total production meets the total demand, maintaining the nominal power grid frequency (50 Hz in Europe). In Denmark, mFRR (as the tertiary reserve) is deployed after the activation of comparatively faster ancillary services when a frequency drop occurs --- mFRR is only used for up-regulation. The mFRR service providers are paid according to their (\textit{i}) capacity in terms of MW, and (\textit{ii}) up-regulation in terms of MWh, if activated. Often, the mFRR providers are not activated, so their upfront earning from the mFRR capacity availability constitutes a passive income. In our setup, we apply a penalty when the mFRR provider fails in the activation stage, i.e., the capacity does not match the actual up-regulation. For the coalition of flexible demands, this could happen due to erroneous flexible capacity estimation or unexpected failures when activating individual assets for up-regulation. An example reason for the activation failure is to avoid the violation of the temperature thresholds of thermostatically controlled loads. 
For further details about mFRR, see \cite{energinet:prequalification}, \cite{energinet:Systemydelser}, and \cite{energinet:tender_conditions_reserves}.

\subsection{Simulation setup for the aggregator}
We use historical mFRR market prices, balancing market prices, and spot (day-ahead) market prices from DK1 bidding zone (western Denmark) in 2022 \cite{energinet:energidataservice}, denoted by $\lambda_{h}^{\text{mFRR}}$, $\lambda_{h}^{\text{b}}$, and $\lambda_{h}^{\text{s}}$, respectively, for every hour $h$. The constant penalty price for the failure in the activation stage is denoted by $\lambda^{\text{p}}$. 

In the reservation stage, e.g., in day $D$-1, we assume that the aggregator considers the current consumption pattern of the coalition as the power baseline $P^{\text{Base}}_{h}$ for the next day and bids capacity $p^{\text{mFRR}}_{h}$ to the mFRR market. This reserve will be activated by the TSO the day after, if needed. In the activation stage, i.e., in day $D$, the true consumption pattern might be different than the baseline, i.e., the consumption pattern of the previous day. Therefore, if the mFRR capacity is activated, indicated by a balancing price higher than the spot price, the coalition may fail in successfully delivering the promised up-regulation (i.e., load reduction). For example, assume in day $D$-1 the coalition has promised a mFRR capacity of 20 kW for a specific hour of the next day. However, in that hour of day $D$, their consummation might be below 20 kW, therefore if activated, the coalition could not deliver the service. The profit of the coalition can be calculated as

\begin{align}\label{eq:mFRRObjective}
     & - \underbrace{\sum_{h=1}^{24} \lambda^{\text{s}}_{h}P^{\text{Base}}_{h}}_{\textrm{Energy cost}} + \underbrace{\sum_{h=1}^{24}\lambda_{h}^{\text{mFRR}} p^{\text{mFRR}}_{h}}_{\textrm{Reservation payment}}  \notag \\ & \quad \quad + \underbrace{\sum_{h=1}^{24}  \lambda_{h}^{\text{b}} p^{\text{b},\uparrow}_{h}}_{\textrm{Activation payment}} - \underbrace{\sum_{h=1}^{24}  \lambda_{h}^{\text{b}} p^{\text{b},\downarrow}_{h}}_{\textrm{Rebound cost}} - \underbrace{ \sum_{h=1}^{24}  \lambda^{\text{p}}s_{h}.}_{\textrm{Penalty cost}}
\end{align}

Accordingly, the profit of the coalition includes the energy cost for purchasing baseline demands $P^{\text{Base}}_{h}$ at spot prices $\lambda^{\text{s}}_{h}$, the reservation payment by selling mFRR services $p^{\text{mFRR}}_{h}$ at mFRR market prices $\lambda_{h}^{\text{mFRR}}$, the activation payment for up-regulations (i.e., load reduction) $p^{\text{b},\uparrow}_{h}$ at balancing prices $\lambda_{h}^{\text{b}}$, the rebound cost by extra consumption $p^{\text{b},\downarrow}_{h}$ at balancing prices $\lambda_{h}^{\text{b}}$, and finally the penalty cost, incurred by the up-regulation not delivered $s_{h}$, penalized at the constant price $\lambda^{\text{p}}$. Note that the first two terms, i.e., the energy cost and the reservation payment, realize in day $D$-1 according to the baseline. However, the remaining three terms are calculated in day $D$ based on the true consumption pattern and the TSO activation.



For the sake of simplicity, we ignore the potential rebound effect of flexible demands by enforcing $p_{h}^{\text{b}, \downarrow} = 0, \ \forall{h}$. In addition, the energy cost is constant irrespective of the flexibility provision, and therefore it can be removed from \eqref{eq:mFRRObjective}. Thus, we rewrite the profit as
\begin{align}\label{eq:mFRR_profit}
     & \underbrace{\sum_{h=1}^{24}\lambda_{h}^{\text{mFRR}} p^{\text{mFRR}}_{h}}_{\textrm{Reservation payment}} + \underbrace{\sum_{h=1}^{24}  \lambda_{h}^{\text{b}} p^{\text{b},\uparrow}_{h}}_{\textrm{Activation payment}} - \underbrace{ \sum_{h=1}^{24}  \lambda^{\text{p}}s_{h}.}_{\textrm{Penalty cost}}
\end{align}


In reality, there is a bidding process for both capacity and up-regulation, however, it is discarded for simplicity. Furthermore, we do not consider the control aspect of the coalition, i.e., the challenge of effectively following the required response within an hour of up-regulation.


In \eqref{eq:mFRR_profit}, $p^{\text{mFRR}}_{h}$, $p^{\text{b},\uparrow}_{h}$, and $s_{h}$ correspond to the whole coalition. Each asset $i$ in the coalition has its own baseline for power consumption, $p_{h, i}$ $\forall{i} \in \mathcal{I}$, such that
\begin{subequations} \label{con}
    \begin{align}
        p^{\text{mFRR}}_{h}        & = \sum_{i} p^{\text{mFRR}}_{h, i}       \\
        p^{\text{b}, \uparrow}_{h} & = \sum_{i}p^{\text{b}, \uparrow}_{h, i} \\
        s_{h}                      & = \sum_{i} s_{h, i}.
    \end{align}
\end{subequations}
%


\subsection{Modeling individual baselines and quantifying the synergy}
For simplicity, we consider homogeneous assets that are supposed to consume 1 kW in one hour of the day, and do not consume any power during the rest of the day. An example for such an asset is a ventilation unit with on/off control. We uniformly distribute assets with 1-hour 1-kW consumption over the day, such that the baseline forecast $p_{h, i}$ for asset $i$ in hour $h$ is defined as
\begin{align}\label{eq:uniform}
    p_{h,i} \thinspace = \begin{cases}
                             1 & \text{if} \quad h  \sim \mathcal{U}(1,24) \\
                             0 & \text{otherwise.}
                         \end{cases}
\end{align}

For an aggregator, it is therefore complex to estimate $p^{\text{mFRR}}_{h, i}$ individually, but as the number of assets increases, the consumption of the coalition converges to a predictable uniform consumption for all $h \in \{1, \hdots, 24 \}$, so $p^{\text{mFRR}}_{h}$ becomes more predictable. Here, the predictability simply means the prediction of the power consumption for the next day, for which more advanced methods can be used \cite{ziras2021baselines}.
This is a stylized example as most assets have some degree of predictability, but it is sufficient to show a relevant problem for coalitions of small-scale uncertain demand-side assets.

Given the profit definition as in \eqref{eq:mFRR_profit}, the synergy effect of the coalition can be quantified as
\begin{align}\label{eq:synergy_effect}
    \text{Synergy effect} = \frac{ \text{Coalition profit} }{ \sum_{i \in \mathbb{I}} \text{Asset \textit{i}'s profit} }
\end{align}
where the synergy effect is simply defined as the ratio between the coalition profit and the sum of individual asset profits. The difference between the two will be due to the activation payment and the penalty cost, as the reservation payments are the same.

\subsection{Shapley values as the payment mechanism}
The Shapley value is a well-known mechanism to allocate contributions in a cooperative game \cite{shapley1997value}. This notion fits perfectly with the concept of synergy effect in a coalition of flexible demands. By using Shapley values, we fairly allocate ex-post the total profit, expressed as in \eqref{eq:mFRR_profit}, collected by the aggregator, among flexible demands.


Let $\mathcal{D}=\{1, \hdots, D \}$ denote the set of flexible demands $d$, where $|\mathcal{D}|$ is the number of flexible demands. Each demand may contain multiple assets, such that $|\mathcal{D}|$ demands altogether have assets $\mathcal{I}=\{1, \hdots, I \}$. Every asset is modeled using \eqref{eq:uniform}.
%
Inspired by \cite{shapley1997value}, the Shapley value $\phi_d$  for the flexible demand $d$, i.e., the payment allocated to that demand, is calculated by
\begin{align}\label{eq:shap}
    \phi_d & = \sum_{\mathcal{S} \subseteq \mathcal{D}, d \in \mathcal{S}} \frac{(|\mathcal{S}|-1)! \ (|\mathcal{D}|-|\mathcal{S}|)!}{|\mathcal{D}|!}\Big[v(\mathcal{S})-v(\mathcal{S} \backslash\{d\})\Big],
\end{align}
where $\mathcal{S}$ represents every possible non-empty subset of the coalition, $|\mathcal{S}|$ is the number of possible subsets,  and $v(\mathcal{S})$ is the value of the coalition subset $\mathcal{S}$, which is the corresponding profit formulated as in \eqref{eq:mFRR_profit}. Similarly, $v(\mathcal{S} \backslash\{d\})$ represents the value of coalition subset $\mathcal{S}$ excluding the flexible demand $d$ with its corresponding assets. The grand coalition is the one including all demands, whose value (profit) is given by $v(\mathcal{D}) = \sum_{d \in \mathcal{D}} \phi_{d}$.


The Shapley value in \eqref{eq:shap} can be thought of as the average marginal contribution of the flexible demand $d$ across all possible coalitions in $\mathcal{D}$. It can only be computed exactly for a relatively small number of flexible demands, as the number of possible non-empty coalitions is equal to $2^{|\mathcal{D}|} - 1$. In this paper, we consider 5 demands, ending up in 31 possible coalitions. The Shapley values can be approximated when $|\mathcal{D}|$ grows \cite{castro2009polynomial}.
Note that the Shapley mechanism has desirable economic properties such as individual rationality and budget balance.


\section{Simulation Results}\label{chapter3}
%

\begin{figure}[!t]
    \centering
    \includegraphics[width=\columnwidth]{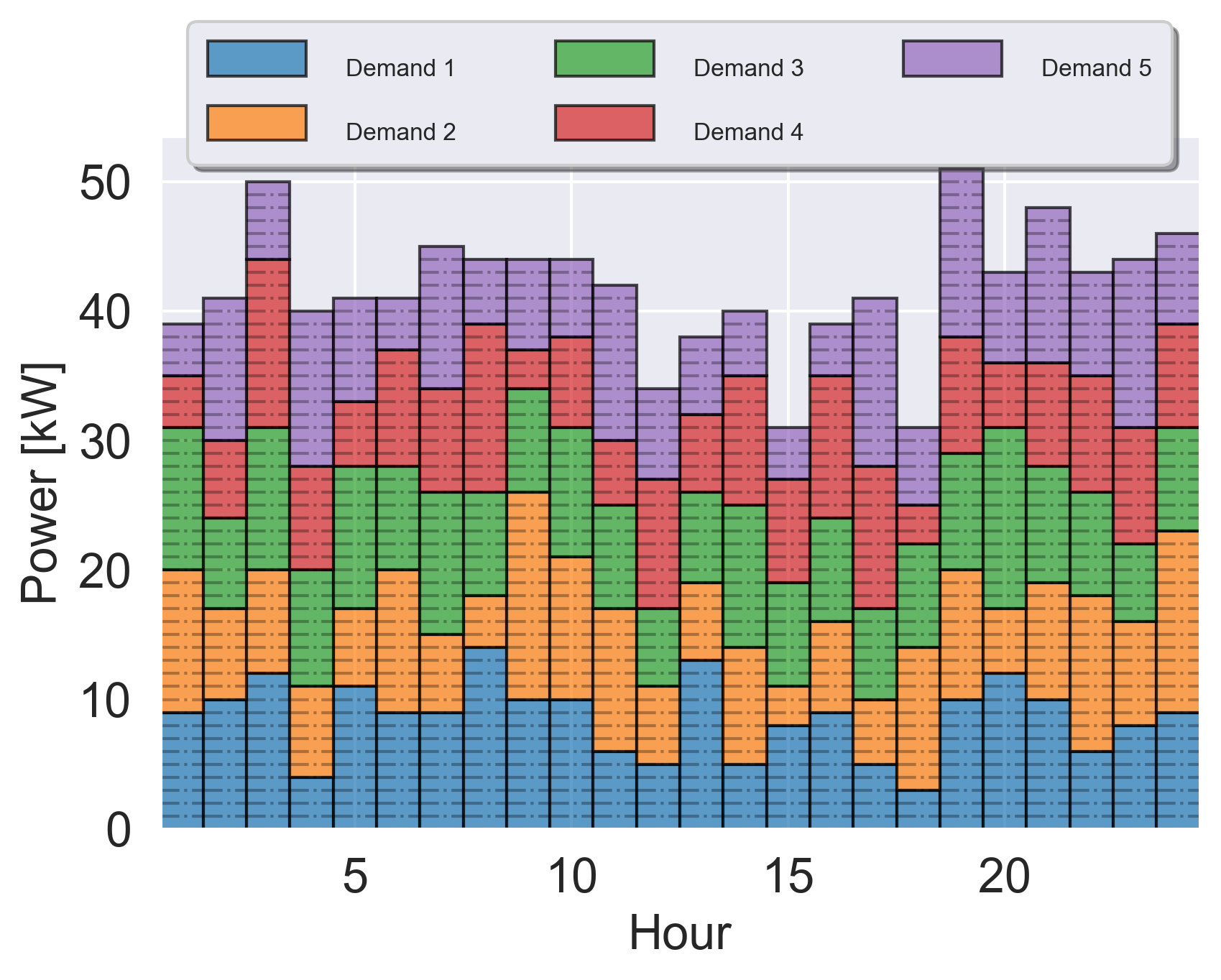}
    \caption{The consumption profile of a coalition with five flexible demands, owning 1000 assets in a sample day. Each asset consumes 1 kW in one hour of the day.}
    \label{fig:assets}
\end{figure}

We consider a coalition of five flexible demands, i.e., $\mathcal{D}=\{d_1, d_2, d_3, d_4, d_5\}$, with 1000 identical assets in total. Each asset consumes 1 kW in one hour of the day, and nothing in the remaining hours. By an on/off control, the consumption reduces from 1 kW to zero. Figure \ref{fig:assets} shows the consumption portfolio for a sample day, where 1000 assets, according to \eqref{eq:uniform}, are uniformly distributed among 24 hours.

We run our simulation over 233 days with the same demands, assets, and ownership, however the daily consumption pattern may differ. As mentioned in Section \ref{chapter3}, the consumption in one day is considered as the baseline for the next day, while the true consumption pattern in the next day might be different. The is the reason that the coalition may fail in successfully delivering the mFRR services, if activated. In particular, this happens if the aggregator has over-promised in the reservation stage, while in the activation stage the consumption level is comparatively lower. We will show in this section that, owed to the synergy effect, it is less likely that the coalition with a comparatively higher number of assets fails to deliver the service.

\begin{figure}[t]
    \centering
    \includegraphics[width=\columnwidth]{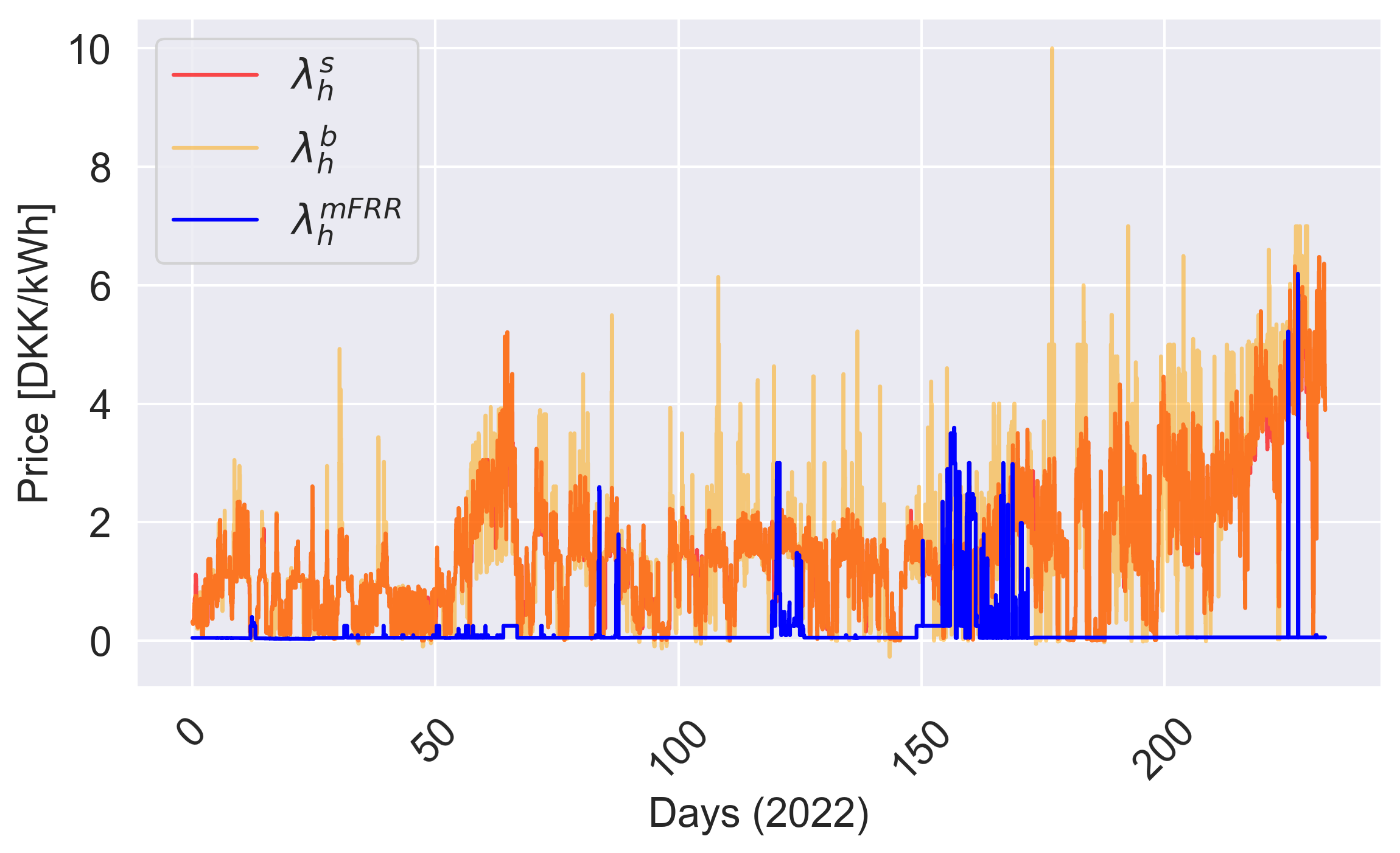}
    \caption{Spot market prices ($\lambda^{\text{s}}_h$), balancing market prices ($\lambda^{\text{b}}_h$), and mFRR market prices ($\lambda^{\text{mFRR}}_h$)  over the first 233 days of 2022 in DK1.}
    \label{fig:prices}
\end{figure}

Figure \ref{fig:prices} illustrates the mFRR market prices $\lambda^{\text{mFRR}}_h$ as well as balancing market prices $\lambda^{\text{b}}_h$ and spot (day-ahead) market prices $\lambda^{\text{s}}_h$ for the first 233 days of 2022 in DK1. The mFRR service is activated by the TSO in every hour $h$ if $\lambda^{\text{b}}_h > \lambda^{\text{s}}_h$.
The penalty price $\lambda^{\text{p}}$ is fixed to 0.1 DKK/kWh. All source codes and input data are
publicly shared in \cite{code}.


\begin{figure}[b]
    \centering
    \includegraphics[width=\columnwidth]{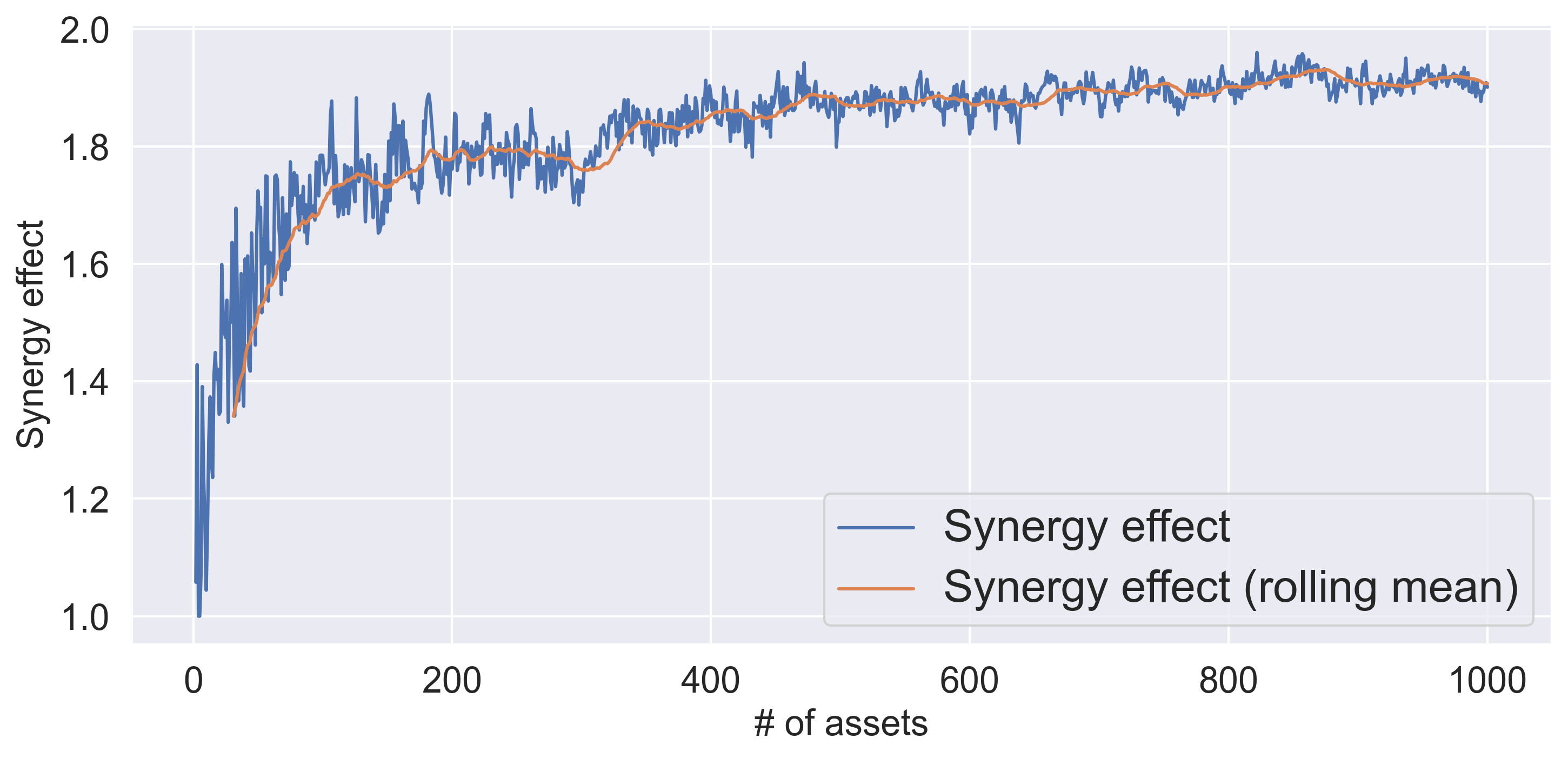}
    \caption{Simulation results: The synergy effect by increasing the number of assets. The rolling mean shows the average of the past 40 point values.}
    \label{fig:synergy_effect}
\end{figure}

Figure \ref{fig:synergy_effect} shows the simulation results, where the synergy effect, as defined in \eqref{eq:synergy_effect}, is quantified for an increasing number of assets, up to 1000. Even with such uncertain assets, it only takes around 400 assets to get the synergy effect curve flattened around 1.9.


\begin{figure}[!t]
    \centering
    \includegraphics[width=\columnwidth]{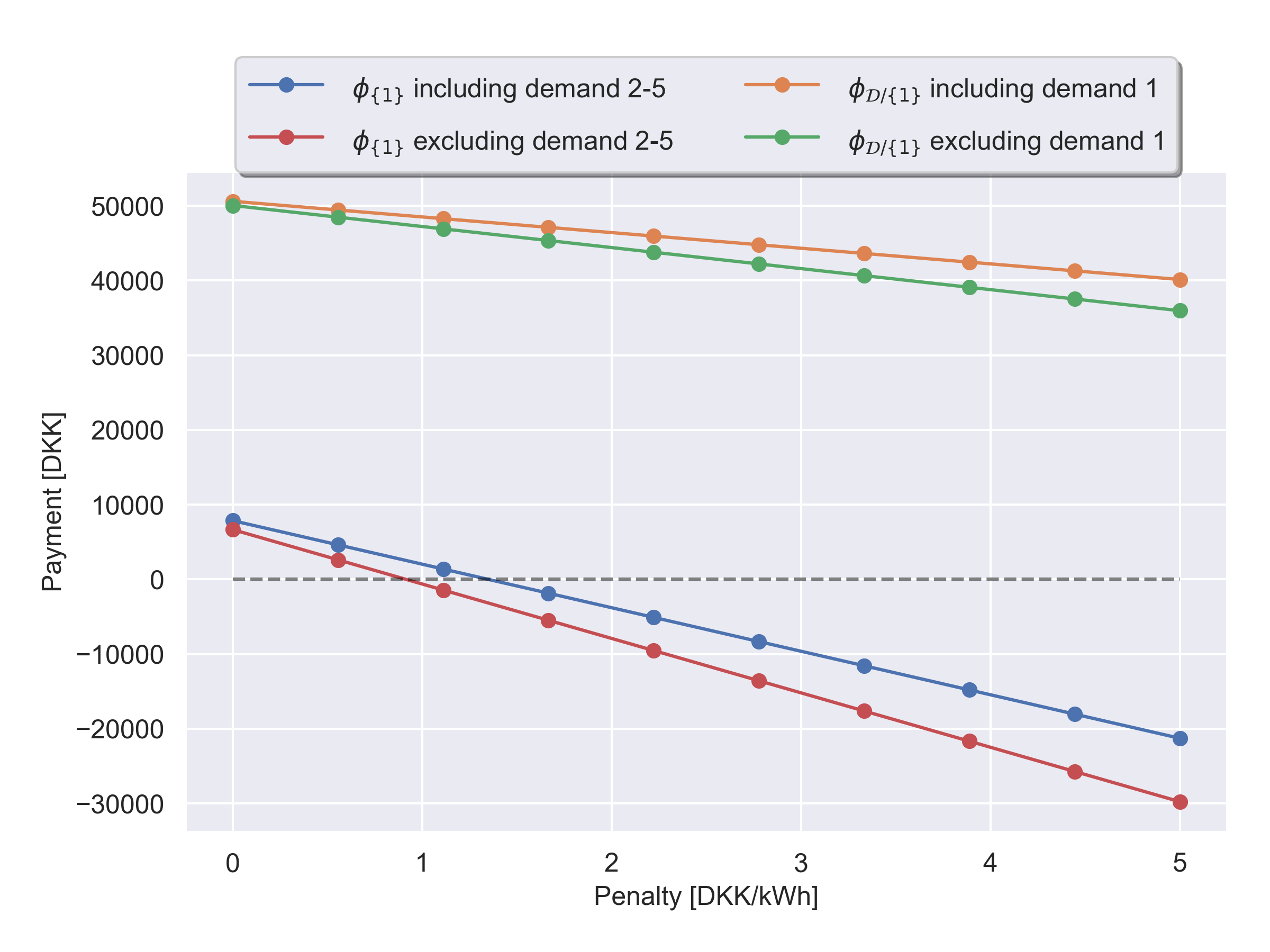}
    \caption{Simulation results: The impact of the penalty price $\lambda^{\text{p}}$ on
        Shapley values (payments $\phi$). The orange curve shows the payment  to $\mathcal{D} / \{d_1\}$, i.e., demands $d_2$ to $d_5$ when $d_1$ is a coalition member. The green curve shows the same payment if $d_1$ was not a member of the coalition. Likewise, the blue (red) curve shows the payment $\phi_{d_1}$ to $d_1$ when it is (it is not) a member of the coalition.}
    \label{fig:shapley_values}
\end{figure}

Figure \ref{fig:shapley_values} shows the Shapley values (payments) in a simulation of an extreme case, wherein the flexible demand $d_1$ always fails in the activation stage. In other words, the assets of $d_1$ never up-regulate and only contribute to the coalition with their reserve capacity (which can never be utilized). We run the simulation for an increasing value of the penalty price $\lambda^{\text{p}}$.
This figure shows that $d_1$ by being within the coalition gets a non-negative payment when $\lambda^{\text{p}} < 1.5$. This is explained by the price data and how often the TSO calls for an activation (recall it happens when $\lambda^{\text{b}} > \lambda^{\text{s}}$). If the power grid would need more up-regulation services, we could expect $d_1$ loses money with penalty prices even lower than 1.5 DKK/kWh. The orange curve shows the payment to $\mathcal{D} / \{d_1\}$, i.e., to the aggregation of four demands $d_2$ to $d_5$, when the coalition includes $d_1$, whereas the green curve shows the same payment when $d_1$ was not a member of the coalition. The difference of these two curves shows that $\mathcal{D} / \{d_1\}$ benefits from having $d_1$ in the coalition, even though $d_1$ gets negative payments. Hence, $\mathcal{D} / \{d_1\}$ is better off with $d_1$ in the coalition.
Similarly, the comparison of blue and red curves shows that $d_1$ also prefers to be part of the coalition by getting paid more (or paying less back). This shows all demands prefer to stay in the coalition.


\section{Conclusion}
\label{chapter4}

This paper illustrated the synergy effect of a coalition of flexible demands, and quantified it in terms of the number of assets bidding to the mFRR market. We also discussed how flexible demands can be fairly paid ex-post using Shapley values. By this, our simulation results showed that all flexible demands are incentivized to stay in the coalition, as opposed to acting on their own. Hence, the synergy effect allows flexible demands to balance the power grid while receiving a revenue for doing so. 

As potential directions for the future work, it is of interest to consider dissimilar assets, treating the heterogeneity as another degree of freedom for further unlocking potential synergy. One could hypothesize that dissimilar assets can further complement each other in terms of flexibility provision. One relevant example could be photovoltaic parks equipped with batteries and/or electrolyzers, providing various frequency services. Moreover, as the number of flexible demands increases, approximate Shapley values must be used due to computational reasons. It is interesting to explore how it impacts the fairness and distribution of payments.

\bibliographystyle{IEEEtran}
\bibliography{bibliography/Bibliography}

\vfill

\end{document}